\documentclass[journal]{IEEEtran}
\usepackage{amsmath, amssymb, bm, cite, epsfig, psfrag}
\usepackage{graphicx}
\usepackage[margin=0.75in]{geometry}
\usepackage{dblfloatfix}
\usepackage{array}
\newcolumntype{P}[1]{>{\centering\hspace{0pt}}p{#1}}
\newcolumntype{M}[1]{>{\centering\hspace{0pt}}m{#1}}
\newcolumntype{L}{>{\centering\arraybackslash}m{3cm}}
\usepackage[font=small]{caption}
\usepackage{epstopdf}	
\usepackage{longtable}
\usepackage{supertabular,booktabs}
\usepackage{multirow}
\usepackage[usenames,dvipsnames]{xcolor}
\renewcommand{\arraystretch}{1.5}
\usepackage{etoolbox}
\usepackage{pbox}
\usepackage{fixltx2e}
\usepackage{filecontents}
\usepackage{enumerate}

\usepackage{colortbl}
\usepackage{fancyhdr}

\usepackage{bm}
\newtoggle{conference}
\togglefalse{conference} 
\graphicspath{{figures/}}

\def\x{\times}

\fancypagestyle{firststyle}{
	\fancyhf{}
	\fancyhead[L]{Y. Xing, O. Kanhere, S. Ju, T. S. Rappaport, and G. R. MacCartney Jr., ``Verification and calibration of antenna cross-polarization discrimination and penetration loss for millimeter wave communications,'' 2018 \textit{IEEE 88th Vehicular Technology Conference (VTC2018-Fall)}, Chicago, USA, Aug. 2018, pp. 1-6.}     

}

\begin{document}
\bibliographystyle{IEEEtran}
\title{Verification and Calibration of Antenna Cross-Polarization Discrimination and Penetration Loss for Millimeter Wave Communications}
\author{\IEEEauthorblockN{Yunchou Xing, Ojas Kanhere, Shihao Ju, Theodore S. Rappaport, and George R. MacCartney Jr.}\\
\IEEEauthorblockA{	\small NYU WIRELESS\\
					NYU Tandon School of Engineering\\
					Brooklyn, NY 11201\\
					\{ ychou, ojask, shao, tsr, gmac\}@nyu.edu}}

\maketitle
\thispagestyle{firststyle}

\begin{abstract}
This article presents measurement guidelines and verification procedures for antenna cross-polarization discrimination (XPD) and penetration loss measurements for millimeter wave (mmWave) channel sounder systems. These techniques are needed to ensure accurate and consistent measurements by different researchers at different frequencies and bandwidths. Measurements at 73 GHz are used to demonstrate and verify the guidelines, and show the consistency of the antenna XPD factor and the penetration loss at different transmitter-receiver (T-R) separation distances, thus providing a systematic method that may be used at any frequency for reliable field measurements.
\end{abstract}
    

\section{Introduction}~\label{sec:intro}

 Recent wireless systems have employed high gain, narrow beamwidth dual-polarized antenna architectures to exploit channel diversity with orthogonally-polarized propagating signals \cite{Ghosh14mmwave}. Furthermore, wideband mmWave networks will require site-specific models that predict the loss induced by common building objects, so that proper 5G and WiFi deployments may be conducted at frequencies far greater than today's IEEE 802.11a networks at 5 GHz. Therefore, ensuring accurate cross-polarization discrimination (XPD) and penetration loss measurement results, and adopting uniform methodologies that may be applied by different institutions at any particular frequency are of necessity to conduct proper and practical network field tests with easy to interpret results. Providing a standard approach to XPD and penetration loss measurements will enable results to be vetted for accuracy and used reliably.
 
 \section{Cross-Polarization Discrimination (XPD)}
 
Characterizing the XPD of antenna systems and radio channels for millimeter wave (mmWave) communication systems using directional antennas is vital for properly interpreting measured results and developing proper path loss models for orthogonally-polarized or dual-polarized communication systems. Even though a transmitted signal may be linearly polarized, scattering effects in the propagation channel will induce some ellipticity to the polarization of the received signal, and the antenna itself may not be ideally linear polarized. Accurate measurement and calibration of the XPD for a directional co-polarized communication system is important for separating antenna and channel effects, where the XPD is defined as the ratio (in dB) of the power in the transmitted co-polarized state to the power radiated in the cross-polarized state when transmitted in free space, without channel impairments \cite{Rap02a,Ippolito2002PropagationEH}. XPD may also be applied to path loss models when determining the received power in co- versus cross-polarized states over distance.

XPD of channels has been studied since the early days of cellphones, in the 1990s.  Measurement results at 1.3 GHz and 4.0 GHz \cite{rappaport92a} showed that the line-of-sight (LOS) channels offered significantly more XPD than the non-line-of-sight (NLOS) channels, and the directional circularly polarized antennas greatly reduced root-mean-square (RMS) delay spread. XPD measurements at 2.6 GHz with 200 MHz bandwidth using a multiple-input-multiple-output (MIMO) channel sounder were presented in \cite{shafi06a}, and models to describe the dependence of XPD on distance, azimuth and elevation and delay spread were investigated, which concluded that the XPD increased with distance and delay. Measurements at 34 GHz with dual-polarized directional horn antennas were conducted to study the behavior of XPD in mmWave channels \cite{iqbal17a}. It was observed that the variation of XPD reduced exponentially with an increase in channel bandwidth. Measurements at 73 \textcolor{black}{GHz} with 800 MHz bandwidth using dual-polarized directional horn antennas \cite{rappaport201573} showed that the XPD was constant over the T-R separation distance range from 10 to 40 m.

\section{Penetration Loss}

 For higher data rates and more reliable links, indoor environments at mmWave need to be extensively investigated for the impact of penetration loss of common building materials, as knowledge of such loss shall be essential to predict indoor and outdoor-to-indoor path loss needed for design and installation of future 5G mmWave wireless systems in and around buildings  \cite{jacque2016indoor,maccartney2015indoor}. Accurate measurements and models for losses induced by partitions, such as walls or floors, will also be important for frequencies well above 100 GHz, as foreseen in future wireless networks. Thus, it is useful to develop a verification methodology that allows researchers to apply a uniform approach to ensure          accurate measurements of partition losses that may be used in site-specific propagation modeling and wireless planning tools.

\section{A standardized verification approach for XPD}\label{sec:XPD}

Using geometric optics and fundamental propagation theory, we have developed verification procedures that may be applied to verify the XPD and penetration loss and are suitable for use as a standard approach. To approve the efficacy of the verification methods, the XPD and penetration loss measurements at 73 GHz were conducted to verify their consistency across various TX and RX antenna types at different TR separation distances. By measuring consistent values over many different distances, relatively close, between the transmitter and receiver, for different frequencies and bandwidths, verification can be performed, ensuring no multipath or antenna artifacts are contained in the measurement system.

The approach validates the XPD of the system antennas. It is repeatable and has been confirmed by measurements at different distances in a controlled, open, and static environment that attempts to remove channel effects and focuses solely on the antennas used. There are three basic rules to follow when measuring the XPD of a transmit and receive antenna for a channel sounder: a) ensure that the measurement is in LOS free space with a T-R separation distance beyond the far-field or Fraunhofer distance $D_f$ of the antennas while also ensuring that the TX and RX antennas are perfectly boresight-aligned; b) ensure that no nearby reflectors or obstructions are present in the propagation path that might cause multipath reflections or induce fading in the measurement; and c) ensure the heights of the antennas and the T-R separation distance between the antennas are selected so that ground bounces and ceiling bounces do not induce reflection, scattering, or diffraction within or just outside the half-power beamwidth (HPBW) of the main lobe of the TX/RX antenna pattern. As shown subsequently, these three rules guarantee accurate measurements of the antenna XPD since the measurement environment is devoid of reflectors or objects that might cause multipath, especially in the first Fresnel-zone which would induce errors into the XPD \cite{Newhall96a, Newhall96b, Newhall97a}.

The three basic rules are further quantified in Fig. \ref{fig:three}:

\begin{figure}    
    \centering
    \includegraphics[width=0.50\textwidth]{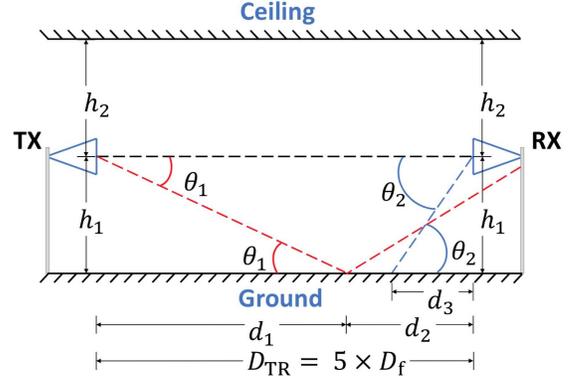}
    \caption{Sketch of geometry and test setup for accurately measuring the antenna XPD between two orthogonally-polarized antennas for channel-sounder verification.}
    \label{fig:three}
\end{figure}

\begin{enumerate}
\item{To ensure that a plane-wave is incident on the RX, measurements in free-space should be made with the TX antenna separated by at least one Fraunhofer distances from the RX antenna. A general rule-of-thumb to assure plane-wave propagation is to set the RX antenna at least five Fraunhofer distances from the radiating TX antenna \cite{Rap02a}. The Fraunhofer far-field distance $D_f$ is defined as:

\begin{equation}\label{equ:Df}
D_f = \dfrac{2D^2}{\lambda},
\end{equation}
where $D$ is the length in meters of the largest linear dimension of the antenna aperture and $\lambda$ is the carrier wavelength of the radiating signal in meters  \cite{Rap02a}. Using the rule of thumb of five Fraunhofer lengths, the T-R separation distance $D_{\text{TR}}$ used to measure the XPD should obey: 

\begin{equation}\label{equ:DTR}
D_{\text{TR}} \geqslant 5 \x D_f.
\end{equation}

Several measurements should be taken at different far-field distances that are greater than  $ 5\times D_f$ and that are far enough from each other to discern an appreciable difference of a few to several dB of received power in free space, while satisfying \eqref{equ:DTR} and the other two rules described below. The additional distances should typically be 20\% to 100\% greater than the initial  distance. For example, if $5\times D_f$ is 4 meters, then 5, 6, and 8 meters would be good distances, as long as they satisfy the other requirements.  Additionally, the TX and RX antennas should be boresight-aligned for both co-polarized and cross-polarized measurements such that their axes of maximum antenna gain align.}

\item {
Following \cite{Newhall96a, Newhall96b, Newhall97a}, the heights of the TX and RX antennas, and the T-R separation distance between the antennas should be chosen so as to avoid any ground, ceiling, wall, or object reflections. Specifically, the heights and distances should be selected in conjunction with the HPBW of the TX and RX antennas such that the projected ground bounce or other reflection sources from the TX antennas are far outside of the HPBW angular spread of the TX antenna and should not arrive anywhere near the HPBW viewing angle of the RX antenna. If the TX antenna has a HPBW of $2\cdot \theta_1$ in radians and the RX antenna has a HPBW of $2\cdot \theta_2$ in radians, and we fix the distance between the TX and RX antennas as $D_{\text{TR}} = 5\times D_{f}$ , then we can use simple geometry to determine the constraint on the height at which the antennas should be placed to avoid multipath sources. Fig. \ref{fig:three} shows a sketch of a typical measurement setup. By solving a set of geometry equations pertaining to the sketch above, the relationship of the T-R separation distance, antenna height above ground ($h_1$) and below the ceiling ($h_2$) can be defined by:

\begin{equation}\label{equ:hh}
h_1, h_2 >\left( \dfrac{D_{\text{TR}}}{\left(\dfrac{1}{\tan(\theta_1)}\right)+\left( \dfrac{1}{\tan(\theta_2)} \right)} \right),
\end{equation}
where $D_{\text{TR}}$ is the T-R separation distance, $h_1$ is the height of the TX and RX antennas above the ground, and $h_2$ is the distance of the antennas from the ceiling and any obstructions or walls on either side of the straight line between the TX and RX antennas. A value twice the height specified in \eqref{equ:hh} is used to ensure additional clearance so as to provide sufficient distance, time, and antenna pattern separation between the direct path and any ground, ceiling, or other reflections in the measurement environment:
\begin{equation}\label{equ:hh2}
h_1, h_2 >2\times \left( \dfrac{D_{\text{TR}}}{\left(\dfrac{1}{\tan(\theta_1)}\right)+\left( \dfrac{1}{\tan(\theta_2)} \right)} \right).
\end{equation}
}

\end{enumerate}

\section{Measurements to Validate XPD values}
A wideband sliding correlator channel sounder with a superheterodyne architecture and directional, high-gain steerable horn antennas was used to conduct the XPD measurements at 73 GHz \cite{Mac17JSACb}. At the TX, a pseudorandom-noise (PN) sequence was generated at the baseband at a transmission rate of 500 Megachips-per-second (Mcps) which was subsequently modulated with a 5.625 GHz intermediate frequency (IF) signal, which was then mixed with a 67.875 GHz local oscillator (LO) to obtain a signal having a center frequency of 73.5 GHz with a 1 GHz radio frequency (RF) null-to-null bandwidth. The 73.5 GHz wideband RF signal was then transmitted through a high-gain and steerable horn antenna. The corresponding super-heterodyne downconverter was employed at the RX with a sliding correlator for baseband processing \cite{Mac17JSACa} (\textcolor{black}{the block diagrams of the channel sounder system are presented in Fig. 1 and 3 of \cite{Mac17JSACa}}).

Two types of horn antennas were used during the XPD measurements: one set of antennas had a 20 dBi gain and a 15 degree HPBW in both azimuth and elevation planes, while the other set had a 27 dBi gain and a 7 degree HPBW in both azimuth and elevation planes. For convenience, we shall henceforth refer to the 15 degree HPBW antennas as widebeam antennas and the 7 degree HPBW antennas as narrowbeam antennas. Three types of TX and RX antenna combinations were used to acquire and verify the XPD values: both the TX and RX antennas were widebeam antennas, both the TX and RX antennas were narrowbeam antennas, and the TX antenna was a widebeam antenna and the RX antenna was a narrowbeam antenna. Henceforth, we shall refer to these antenna combinations as ``wide-to-wide'', ``narrow-to-narrow'' and ``wide-to-narrow'', respectively. The antennas could be set in a cross-polarized configuration by rotating one by 90 degree relative to the other and with the same direction of transmission or reception, in a systematic way either electrically or mechanically to make both antennas experience identical cross-polarization from each other. During the measurements, the TX antenna was kept vertically polarized and the RX antenna was either vertically or horizontally polarized. 

The largest linear dimension of the horn antenna aperture was 0.02 m for the widebeam antennas and 0.041 m for the narrowbeam antennas. Therefore, at 73.5 GHz, the Fraunhofer distances $D_f$ were 0.196 m and 0.824 m for the widebeam and narrowbeam antennas, respectively. Following \eqref{equ:DTR}, 0.98 m and 4.12 m were the corresponding minimum T-R separation distances for widebeam and narrowbeam antennas. By following \eqref{equ:hh2} the minimum heights of the TX/RX antennas were 1.34 m for the widebeam antennas at $D_{\text{TR}} = 5$  m and 0.61 m for the narrowbeam antennas at $D_{\text{TR}} = 5$  m.

Following these rules \eqref{equ:DTR} \eqref{equ:hh2}, T-R separation distances of 3, 3.5, 4, 4.5, and 5 m were used, and TX/RX antenna heights of 1.5 m were chosen.










\section{XPD Analysis}
The XPD is calculated by first measuring the path loss between the co-polarized TX and RX antennas at several different distances. With antenna gains removed, the path loss is presented as:

\begin{equation}\label{equ:PLVV}
\begin{split}
PL_{\text{V-V}} (d) (\text{dB}) = P_\text{{t-V}}(\text{dBm}) - P_\text{{r-V}}(d)(\text{dBm}) \\+ G_\text{{TX}}(\text{dBi}) + G_\text{{RX}}(\text{dBi}),
\end{split}
\end{equation}
where $d$ is the T-R separation distance in meters, $P_\text{{t-V}}$ is the transmitted power into the vertically polarized TX antenna in dBm, $P_\text{{r-V}}(d)$  is the received power at the output of the vertically polarized RX antenna in dBm at a distance $d$, $G_\text{{TX}}$ is the gain of the TX antenna in dBi, $G_\text{{RX}}$ is the gain of the RX antenna in dBi, and $PL_{\text{V-V}}(d)$ is the measured path loss in dB at a distance $d$. Note that antenna gains are not necessary to calculate the XPD, but are required for accurately measuring and calibrating far-field free space path loss. 

The cross-polarized path loss is then calculated at the same distance with the cross-polarized antennas (i.e., V-H) as follows:

\begin{equation}\label{equ:PLVH}
\begin{split}
PL_{\text{V-H}} (d) (\text{dB}) = P_\text{{t-V}}(\text{dBm}) - P_\text{{r-H}}(d)(\text{dBm}) \\+ G_\text{{TX}}(\text{dBi}) + G_\text{{RX}}(\text{dBi}),
\end{split}
\end{equation}
where the $P_{t-V}$, $G_{TX}$, and $G_{RX}$ are defined as above, $P_\text{{r-H}}(d)$ is the received power in dBm at the output of the horizontally-polarized RX antenna at a distance $d$, and $PL_{\text{V-H}}(d)$  is the cross-polarized path loss in dB at a distance $d$. The XPD between the antennas in dB is then found by subtracting $PL_{\text{V-V}}(d)$(dB) from $P_\text{{r-H}}(d)$(dB) at different T-R separation distances:

\begin{equation}
XPD(d)(\text{dB}) = PL_{\text{V-H}}(d)(\text{dB}) - PL_\text{{V-V}}(d)(\text{dB}),
\end{equation}
where $XPD(d)$ is typically a positive value in dB. Since the T-R separation distances are identical for the co- and cross-polarized FSPL measurements at a particular location, the difference between the two values may be considered to be the XPD between the arriving signals, induced by the differences in antenna polarization at the TX and RX. A number of measurements following the procedure outlined above should be made at various distances in the far-field to ensure measurement accuracy.

\section{XPD Measurement Results}
To make sure that there are no additional multipath or scatterers used to calculate the XPD values, only the received power of the first resolvable arriving multipath component is used for each PDP for both co-polarized and cross-polarized antenna combinations. The three guidelines ensure that there are no multipath components within the first arriving component.

\begin{figure}    
    \centering
    \includegraphics[width=0.50\textwidth]{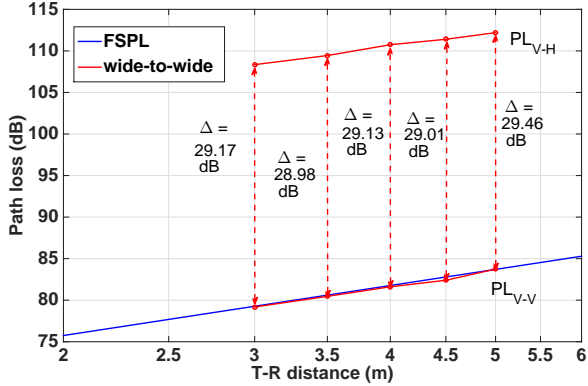}
    \caption{XPD of wide-to-wide antenna combination, $PL_{V-V}$ is the measured path loss using co-polarized antennas, and $PL_{V-H}$ is the measured path loss using the cross-polarized antennas. \textcolor{black}{The antenna heights of the both TX and RX are 1.5 m.}}
    \label{fig:XPDww}
\end{figure}

Table \ref{tbl:XPD} provides the results of the XPD measurements for the three antenna combinations at 3 m, 3.5 m, 4 m, 4.5 m and 5 m. See Fig. \ref{fig:XPDww} for the XPD values measured at all of the five distances for wide-to-wide antenna combination. It is clear that the measurements are all within 0.5 dB of each other, which demonstrates that XPD is constant and independent of the T-R separation distance for far-field propagation. It is worth noting that the mean XPD values of the three antenna combinations are within 1 dB of each other over all distances, which indicates that the antenna combination does not have a significant effect on the XPD -- an important expected result that should occur during verification.

\begin{table*}[!ht]\footnotesize
\renewcommand{\arraystretch}{1.1}
\centering
\caption{XPD Measurement Results at 73 GHz} \label{tbl:XPD}
\newcommand{\tabincell}[2]{\begin{tabular}{@{}#1@{}}#2\end{tabular}}
\begin{tabular}{|c|c|c|c|c|c|c|c|}
\hline
\tabincell{c}{ }&\tabincell{c}{\textbf{3 m}}&\tabincell{c}{\textbf{3.5 m}}&\tabincell{c}{\textbf{4 m}}&\tabincell{c}{\textbf{4.5 m}}&\tabincell{c}{\textbf{5 m}}&\tabincell{c}{\textbf{Mean}}&\tabincell{c}{\textbf{$\sigma$}}\\
\hline
\tabincell{c}{Wide-to-Wide XPD }&\tabincell{c}{{29.17 dB}}&\tabincell{c}{{28.98 dB}}&\tabincell{c}{{29.13 dB}}&\tabincell{c}{{29.01 dB}}&\tabincell{c}{{29.46 dB}}&\tabincell{c}{{29.15 dB}}&\tabincell{c}{{0.19 dB}}\\
\hline
\tabincell{c}{Narrow-to-Wide XPD }&\tabincell{c}{{28.73 dB}}&\tabincell{c}{{28.98 dB}}&\tabincell{c}{{29.58 dB}}&\tabincell{c}{29.42 dB}&\tabincell{c}{29.79 dB}&\tabincell{c}{{29.30 dB}}&\tabincell{c}{{0.44 dB}}\\
\hline
\tabincell{c}{Narrow-to-Narrow XPD }&\tabincell{c}{28.54 dB}&\tabincell{c}{30.60 dB}&\tabincell{c}{30.17 dB}&\tabincell{c}{30.9 dB}&\tabincell{c}{31.31 dB}&\tabincell{c}{{30.30 dB}}&\tabincell{c}{{1.07 dB}}\\
\hline
\end{tabular}
\end{table*}


\section{A standardized approach for Penetration Measurement Verification}

The standardized procedure for verifying the accuracy of measuring material penetration loss with a channel sounder described here is nearly identical to the procedure for verifying and measuring antenna XPD values in Section \ref{sec:XPD}, but with a few extra requirements. Calculation of penetration loss is done by first measuring the received power in free-space for several distances in the far field of both the TX and RX antennas for vertical-to-vertical, vertical-to-horizontal, horizontal-to-vertical, or horizontal-to-horizontal polarized antennas. Then, the additional loss caused by a material under test (MUT) half-way between the TX and RX antennas is measured \cite{jacque2016indoor}, and then the measurement system is brought in-situ across the partition to be measured \cite{jacque2016indoor}. Multiple separation distances for both the free space and partition-separated measurements are needed to increase accuracy and guarantee validity of the measurement. It can be verified that material penetration loss measurements performed for a given frequency and material are accurate if the material loss does not vary as a function of the distance between the TX and RX antennas.

\begin{figure}    
    \centering
    \includegraphics[width=0.45\textwidth]{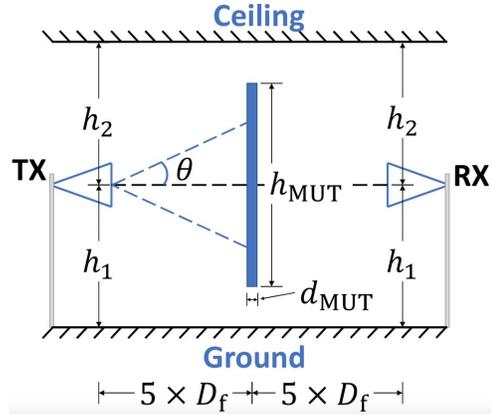}
    \caption{Sketch of geometry and test setup requirements for accurately measuring material penetration loss for channel sounder calibration and verification.}
    \label{fig:MUT}
\end{figure}

The following requirements are necessary for accurate material penetration loss measurements:

\begin{enumerate}
\item{To ensure that a plane wave is incident upon the MUT, the material should be placed at least five Fraunhofer distances \eqref{equ:Df} from the radiating TX antenna. The RX antenna should be at least one Fraunhofer distance on the other side of the MUT, but for consistency, it should also be five Fraunhofer distances from the MUT, if possible. Thus, the T-R separation distances used to measure penetration loss should obey: 
	
\begin{equation}\label{equ:DTR2}
D_{\text{TR}} \geqslant 10 \x D_f.
\end{equation}
}

\item The TX and RX antennas should be equidistant from the center of the MUT. Typically, the MUT  should be oriented perpendicularly to the direction of propagation. The propagating planar-wave front should illuminate a large cross-section of the MUT, and only the MUT. The TX and RX antennas must be boresight-aligned during both the free-space and the MUT measurements.


\item {The MUT should have dimensions large enough such that the radiating wavefront from the TX antenna is illuminated on the MUT without exceeding the projected HPBW angle spread from the TX antenna. That is to say that the MUT should have a larger surface area than the cross-section illumination of the TX antennas' HPBW spread at the distance to the MUT. We note that omnidirectional antennas may be used to measure penetration loss, but the channel sounder used should have extremely-fine temporal resolution in order resolve individual multipath components for analysis so that only the component that penetrates the MUT is measured (more details on bandwidth/time resolution are described below). Fig. \ref{fig:MUT} displays a side view (or top-down view) of the cross-section and the illumination requirements for material penetration loss measurements. The height and width of the MUT must be considerably larger than the length defined by the geometry of the antenna HPBW and the distance between the TX antenna and the material. For a TX antenna with a given HPBW in radians and distance in meters from the MUT, the height and width of the MUT in meters should be:

\begin{equation}\label{equ:MUT}
h_{MUT} >>5\times 2 \times D_f \cdot \tan\left( \dfrac{HPBW}{2} \right),
\end{equation}
where the $HPBW$ is the transmit antenna HPBW in radians, $D_f$ is the Fraunhofer far-field distance in meters, and $h_{MUT}$ is the height or width (from edge-to-edge) of the MUT in meters. 
}

\item The measurement environment should be devoid of reflectors or objects (aside from the MUT) that might cause multipath, which could induce errors. More specifically, multipath components in the PDP measurement should be 20-30 dB down from the main arriving path, or at resolvable time-delay bins that are separable from the direct path between the TX and RX antennas such that they can be removed in post-processing.

\item The heights of the antennas and the T-R separation distance should be chosen so as to avoid ground or ceiling bounce reflections. Specifically, the heights and distances should be selected in conjunction with the HPBWs of the TX and RX antennas such that the projected ground bounce, ceiling bounce, or side obstructions from the TX antenna HPBW spread do not arrive within the HPBW viewing angle of the RX antenna. Note that extremely large MUTs or partitions will by nature not result in ground bounces or ceiling bounces, especially at mmWave frequencies.

\item While narrowband and wideband channel-sounder operation can be used for measuring penetration loss, there are materials and conditions for which multiple internal reflections occur within the MUT \cite{Du2016a}. Such internal reflections can cause multipath, and constructive and destructive interference that contributes to penetration loss measurements as observed at the RX. Therefore, it is suggested that penetration loss measurements be conducted with an extremely wideband channel sounder in order to resolve multipath components such that they are separable in time-delay bin from the direct path. This is particularly important when measuring materials with internal structures that cause internal reflections and multipath components that travel closely in time with the direct path. By measuring penetration loss with a wideband channel sounder, the power in the first arriving multipath component at the RX may be used to compute the penetration loss between the free-space measurement and the MUT measurement \cite{jacque2016indoor}.

\end{enumerate}

\section{Penetration loss measurements Analysis and results}

\begin{table*}[!ht]
\renewcommand{\arraystretch}{1.1}
\centering
\caption{Penetration Loss Measurement Results at 73 GHz} \label{tbl:MUT}
\newcommand{\tabincell}[2]{\begin{tabular}{@{}#1@{}}#2\end{tabular}}
\begin{tabular}{|c|c|c|c|c|c|}
\hline
\tabincell{c}{MUT\\Clear Glass (1.2 cm thickness) }&\tabincell{c}{\textbf{3 m}}&\tabincell{c}{\textbf{4 m}}&\tabincell{c}{\textbf{5m}}&\tabincell{c}{\textbf{Mean}}&\tabincell{c}{\textbf{$\sigma$}}\\
\hline
\tabincell{c}{V-V}&\tabincell{c}{{7.54 dB (6.28 dB/cm)}}&\tabincell{c}{{7.39 dB (6.15 dB/cm)}}&\tabincell{c}{{8.23 dB (6.86 dB/cm)}}&\tabincell{c}{{7.72 dB (6.43 dB/cm)}}&\tabincell{c}{{0.45 dB }}\\
\hline
\tabincell{c}{V-H}&\tabincell{c}{8.48 dB (7.06 dB/cm)}&\tabincell{c}{7.16 dB (5.96 dB/cm)}&\tabincell{c}{7.62 dB (6.35 dB/cm)}&\tabincell{c}{{7.75 dB (6.46 dB/cm)}}&\tabincell{c}{{0.67 dB}}\\
\hline
\end{tabular}
\end{table*}

Penetration measurements were conducted at the NYU WIRELESS research center, where T-R separation distances of 3 m, 4 m, and 5 m were used according to~\eqref{equ:DTR2}, the TX/RX antenna heights were 1.5 m which satisfied \eqref{equ:hh2}, and the straight line drawn between the TX and RX antennas should pass through the center of the MUT. The wide-to-wide antenna combination for which the XPD was calculated was used to measure the penetration loss of the MUT -- both co- and cross-polarized penetration losses were measured and calculated as:

\begin{equation}\label{equ:LVV}
\begin{split}
L_{V-V}\text{[dB]} = P_t\text{[dBm]}-P_{r-V}^{MUT}(d)\text{[dBm]} + G_{TX}\text{[dBi]}\\
+G_{RX}\text{[dBi]}-PL_{V-V}(d)\text{[dB]},
\end{split}
\end{equation}

\begin{equation}\label{equ:LVH}
\begin{split}
L_{V-H}\text{[dB]} = P_t\text{[dBm]}-P_{r-H}^{MUT}(d)\text{[dBm]} + G_{TX}\text{[dBi]}\\
+G_{RX}\text{[dBi]}-PL_{V-H}(d)\text{[dB]}-XPD\text{[dB]},
\end{split}
\end{equation}
where $P_{r-V}^{MUT}(d)$ and $P_{r-H}^{MUT}(d)$  are the co- and cross-polarized received powers in dBm, respectively, at distance $d$ in meters at the output of the RX antenna with the MUT between the TX and RX antenna, and $L_{V-V}\text{[dB]}$ and $L_{V-H}\text{[dB]}$ are the co- and cross-polarized material penetration loss.

The normalized penetration loss of the materials was also measured in dB/cm, which can subsequently be used to find the average normalized penetration loss for an ensemble of commonly measured materials of different thicknesses such as glass and drywall as was done in \cite{jacque2016indoor}.

\begin{figure}    
    \centering
    \includegraphics[width=0.50\textwidth]{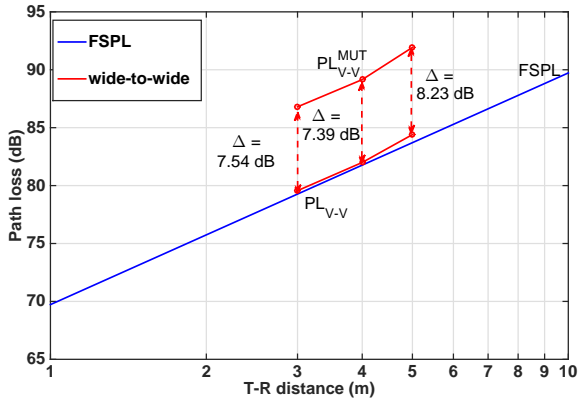}
    \caption{Penetration loss of clear glass, $PL_{V-V}$ is the measured path loss using co-polarized antennas, and $PL_{V-V}^{MUT}$ is the measured path loss with the MUT in the path of propagation.}
    \label{fig:Pww}
\end{figure}

The penetration loss measurement results are shown in Table \ref{tbl:MUT} and Fig. \ref{fig:Pww}. For both co-polarized and cross-polarized antennas, the penetration loss does not vary significantly over the T-R separation distances measured. The standard deviation $\sigma$ of the penetration loss measured over distance for both co-polarized and cross-polarized antennas is less than 1 dB. The average penetration loss in both cases is approximately 7.7 dB (6.43 dB/cm).

\section{Conclusion}
A universal standard approach that can verify proper calibration and measurements of XPD and penetration loss for mmWave signals was described in this paper. Measurements at 73 GHz showed that when following the setup given in Fig. \ref{fig:three}, and equations \eqref{equ:DTR} and \eqref{equ:hh2}, XPD values were independent of the T-R separation distances and the antenna combinations when in the far field but relatively closely separated (within 4-8 m). A standard approach for performing penetration loss measurements was given in Fig. \ref{fig:MUT}, and equations \eqref{equ:DTR2} and \eqref{equ:MUT}, and measurements at 73 GHz for V-V and V-H polarization configurations were presented for clear glass. It was shown that  the penetration loss was constant over T-R separation distances, as verified by the stable 7.7 dB penetration loss for clear glass at 73 GHz at different distances for both V-V and V-H polarizations.

\section{Acknowledgment}
This work is supported in part by the NYU WIRELESS Industrial Affiliates, and in part by the GAANN Fellowship Program, and in part by the National Science Foundation under Grants: 1320472, 1237821, 1302336, 1702967 and 1731290. Thanks to Kate Remley of NIST for her suggestions and help during the discussions.

\bibliographystyle{IEEEtran}
\bibliography{XPD}

\end{document}